\documentclass[prd,aps,superscriptaddress,twocolumn,floatfix,nofootinbib]{revtex4-1}
\pdfoutput=1

\usepackage{amsfonts}
\usepackage{amsmath}
\usepackage{amssymb}
\usepackage{bm}
\usepackage{dcolumn}
\usepackage{graphicx}   
\usepackage[latin1]{inputenc}
\usepackage{latexsym}
\usepackage{rotating}
\usepackage{hyperref}
\usepackage{graphicx}
\usepackage{color}

\newcommand\be{\begin{equation}}
\newcommand\ba{\begin{eqnarray}}
\newcommand\ee{\end{equation}}
\newcommand\ea{\end{eqnarray}}

\begin{document}

\title{Superstring Cosmology - A Complementary Review}

\author{Robert Brandenberger}
\email{rhb@physics.mcgill.ca}
\affiliation{Department of Physics, McGill University, Montr\'{e}al,
  QC, H3A 2T8, Canada}
 

\begin{abstract}

In this review,  a number of approaches to superstring cosmology which make use of key features which distinguish string theory from point particle theories are discussed,  with particular emphasis on {\it emergent} scenarios.  One motivation for the discussion is the realization that,  in order to describe the evolution of the very early universe, it is necessary to go beyond a conventional effective field theory (EFT) analysis.   Some of the conceptual problems of an EFT analysis will be discussed. The review begins with a summary of the criteria for a successful early universe scenario,  emphasizing that cosmic inflation is not the only scenario of early universe cosmology which is consistent with current cosmological observations.  Bouncing and emergent scenarios as interesting alternatives are introduced.  Some realizations of these scenarios from superstring theory are reviewed, e.g. String Gas Cosmology,  the Pre-Big-Bang scenario,  the Ekpyrotic model, Double Field Theory cosmology and matrix model cosmology.  In light of the difficulties in obtaining cosmic inflation from string theory (at the level of EFT), and realizing that there are promising examples of alternative early universe scenarios which are derived from basic principles of superstring theory,  one must entertain the possibility that the cosmology emerging from string theory will not involve an extended period of accelerated expansion.  

\end{abstract}

\maketitle

\section{Introduction} 
\label{sec:intro}

Recently, a comprehensive review of superstring cosmology appeared \cite{Review}. Specifically, the review focused on string moduli and their stabilization, on attempts to obtain inflation in string theory, and on connections to late time cosmology.  Most of the discussions were based on (point particle-based) effective field theory models. But these are unable to take into account a number of key differences between string theory and point particle theories and hence may be missing crucial insights.  The main motivation to write this short complementary review is to highlight these differences and discuss how the cosmology emerging from superstring theory may be quite different from what one would expect using effective field theory models. The focus of this review will be on approaches which the author has worked on.
 
This review starts by explaining the key requirements for an early universe cosmology to be consistent with current data. Inflation is one scenario which satisfies these requirements, but it is not the only one.  In Section \ref{section2} I will discuss how bouncing and emergent cosmologies can also provide successful early universe scenarios.

To obtain bouncing and emergent scenarios, it is necessary to go beyond an effective field theory description by which I will mean a model based on Einstein gravity as a theory of space and time coupled to matter fields obeying the null energy condition. In  Section \ref{section3} I will argue that effective field theory descriptions of accelerating cosmologies are subjec to a number of conceptual problems, some of which follow from superstring theory while others are more general. The bottom line of this discussion is that we need to go beyond an effective field theory description if we wish to understand the evolution of the early universe at the string density.
 
 One of the problems of the standard effective field theory analysis is that key symmetries of string theory are hidden. In Section \ref{section4} I will discuss a couple of approaches which aim to include these symmetries while maintaining a desription which is based on fields.  Specifically, I will mention the {\it Pre-Big-Bang Scenario} \cite{PBB}, the Ekpyrotic model \cite{Ekp} and {\it Double Field Theory} \cite{DFT} cosmology.  These approaches, however, focus on the massless modes of string theory, and at the high densities of the early universe it should be expected that the full tower of string modes becomes important.  Thus, in the rest of this review we turn to non-perturbative approaches. First, (section \ref{section5}) we review {\it String Gas Cosmology} (SGC), a scenario which is based of incorporating the key new degrees of freedom and new symmetries which differentiate string theory from point particle theories \cite{BV}.  SGC is a toy model and is built upon a space-time background which is not consistently derived from string theory. In section \ref{section6} we explore recent attempts to derive early universe cosmology from truly non-perturbative approaches to superstring theory \cite{BBL}.
 
 We work in natural coordinates in which the speed of light, Planck's constant and Boltzmann's constant are all set to $1$. In the first three sections of this article we work in the context of a spatially flat Friedmann-Lemaitre-Robertson-Walker (FLRW) metric
 \be
 ds^2 \, = \, dt^2 - a(t)^3 d{\bf{x}}^2 \, ,
 \ee
 where $t$ is physical time, ${\bf{x}}$ are comoving spatial coordinates, and $a(t)$ is the scale factor which measures the physical radius of spatial sections. It is convenient to normalize $a(t)$ to be $a(t_0) = 1$ at the present time $t_0$.  The time of recombination (when the cosmic microwave background (CMB) is released) is denoted by $t_{rec}$. The temperature at time $t$ is denoted by $T(t)$.  
 
 If we work in the context of Einstein's theory of space and time coupled to matter with energy density $\rho$ and pressure $p$, then the equations of motion for the geometry are
 \be \label{FRW1}
 H^2 \, = \, \frac{8 \pi} G \rho \, ,
 \ee
 where $H(t)$ is the expansion rate of space at time $t$ and $G$ is Newton's gravitational constant,   and
 \be \label{FRW2}
 {\dot H} \, = \, - 4 \pi G (\rho  + p) \, ,
 \ee
an overdot denoting the derivative with respect to time.
 
 It is very important to distinguish two key length scales in cosmology. The first is the {\it horizon}, which is the forward light cone of a point on some initial Cauchy surface, and which carries information about causality. The second key scale is the Hubble radius
 \be
 l_H(t) \, = \, H^{-1}(t) \, .
 \ee
As discussed in detail in \cite{MFB} (see also \cite{RHBfluctsrev} for a shorter overview),  linear fluctuations about a  FLRW background are frozen on length scales greater than the Hubble radius, while they oscillate on sub-Hubble scales.  Thus, the Hubble scale is key to understand the generation and evolution of cosmological fluctuations.
 
 Cosmological fluctuations play a crucial role in cosmology.  These should be visualized as a superposition of small amplitude plane wave perturbations of density and pressure. Since on large cosmological scales the fluctuations have a small amplitude today and since gravity is an attractive force, these perturbations were even smaller in the early universe. Their evolution can hence be described by linear perturbation theory. In this approximations, all comoving Fourier modes evolve independently (their physical length scales with $a(t)$).  Fluctuations generated in the early universe propagate to the present time and provide us a window to probe high energy scale physics with current cosmological observations. 
 
\section{Early Universe Scenarios}
\label{section2}

The key challenge for early universe cosmology is to provide a scenario which explains the observed structure of the universe. First, the scenario must produce a universe which is nearly spatially flat and isotropic today.  Second, there needs to be a causal mechanism for the generation of small amplitude inhomogeneities which can explain the observed angular power spectrum of the CMB (see Fig. 1) which is characterized by a flat spectrum on the largest angular scales, and the famous acoustic oscillations on smaller scales, with a first peak at the angular scale which corresponds to the comoving Hubble radius at the time of recombination.  

\begin{center}
\begin{figure}[!htb]
\includegraphics[width=6.4cm]{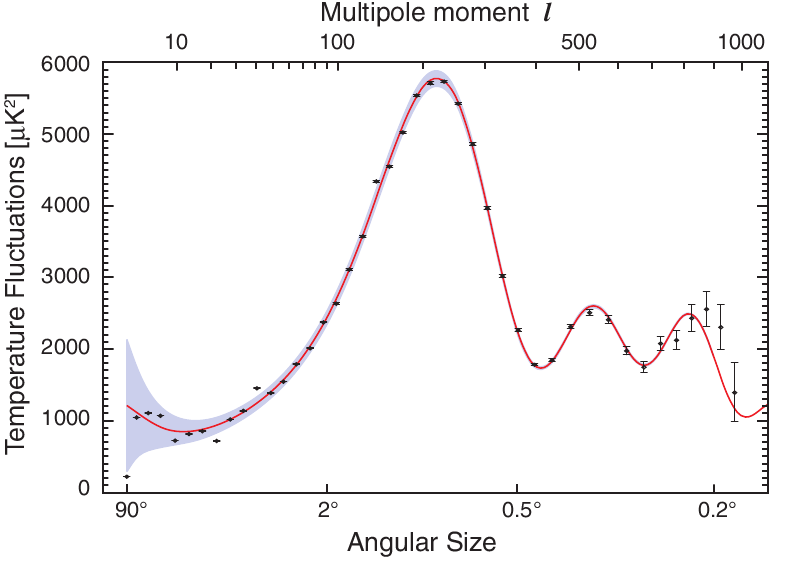}
\caption{Angular power spectrum of the CMB (WMAP results \cite{WMAP}). The horizontal axis indicates the angular scale, the vertical axis is the power of anisotropies on the corresponding scale. Credit: NASA/WMAP Science Team.}
  \label{fig1}
\end{figure}
\end{center}

Note that physics which can yield these acoustic oscillations in the angular power spectrum of the CMB was in fact discussed in detail in two classic works \cite{SZ, PY} ten years before the devolopment of inflationary cosmology. It was assumed that in the early universe there was a nearly scale-invariant spectrum of cosmological fluctuations on length scales which extended beyond the Hubble radius at the time of recombination. One should picture these fluctuations as standing waves which begin to oscillate once their wavelength drops below the Hubble scale. The scale with wavenumber $k_1$ which enters the Hubble radius at $t_{rec}$ has not had time to oscillate.  Hence, the surface of last scattering on this length scale has a fluctuation of maximal amplitude at this scale, leading to a peak in the CMB anisotropies.  The mode with wavenumber $k_2$ which entered the Hubble radius slightly earlier and has had time to do a quarter of an oscillation between the time when it enters and the time of last scattering is caught at a node in the fluctuation amplitude, leading to a local minimum of the angular power spectrum.  This is illustrated in Fig.  2, a sketch redrawn from \cite{SZ}.

\begin{center}
\begin{figure}[!htb]
\includegraphics[width=6.4cm]{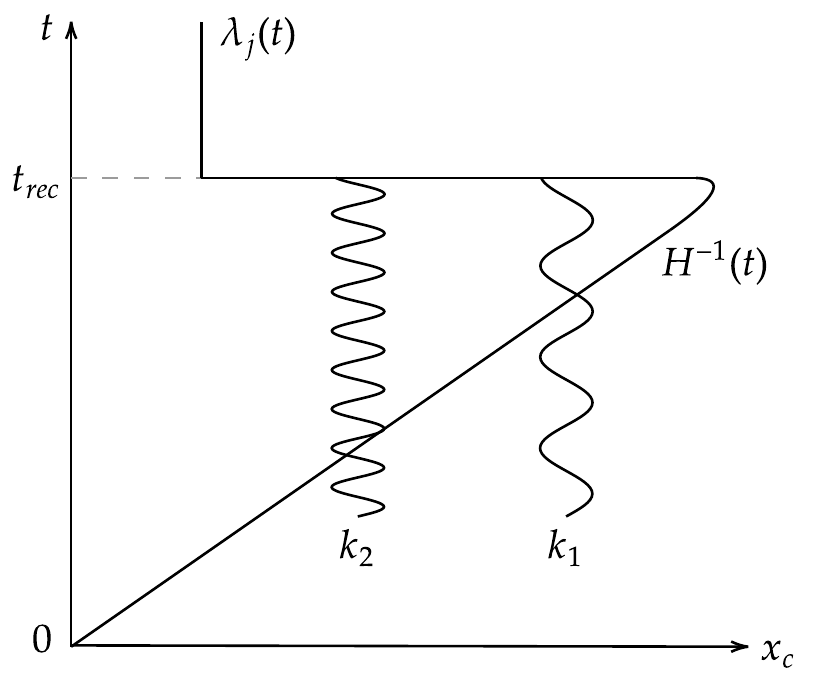}
\caption{Space-time sketch (redrawn based on an original figure in \cite{SZ}) illustrating the origin of the acoustic oscillations in the CMB angular power spectrum. The vertical axis is time, with $t_{rec}$ denoting the time of recombination when the microwave background radiation is released. The horizontal axis indicates comoving spatial scale.  Fluctuations on two different length scales are denoted by $k_1$ and $k_2$. Fluctuations enter the Hubble horizon $H^{-1}(t)$ as standing waves and begin to oscillate. The mode which enters at $t_{rec}$ has not had time to oscillate and has maximal amplitude at $t_{rec}$, when the mode indicated by $k_1$ has undergone $5/4$ oscillations and is at a node at $t_{rec}$.  The post-recombination Jeans length is indicated by the line $\lambda_J(t)$.}
  \label{fig-SZ}
\end{figure}
\end{center}

The origin of an approximately scale-invariant spectrum of cosmological fluctuations was not discussed in the original works. Since in Standard Big Bang cosmology the Hubble radius and the horizon coincide (up to a factor of order unity), super-Hubble fluctuations in fact appear to be acausal.  A successful early universe scenario must hence obey the following criteria:\\
1.  In order to explain the near isotropy of the CMB, the horizon at $t_{rec}$ must be several orders of magnitude larger than the Hubble radius.\\
2. To obtain a causal mechanism of structure formation, the physical length of fluctuation modes observed today must originate inside the Hubble radius at some early time.\\
3. There must be a causal generation mechanism which produces a spectrum of nearly scale-invariant cosmological fluctuations with a slight red tilt.

Cosmological inflation \cite{Guth} was the first  scenario proposed in which the three above criteria can be satisfied. The scenario assumes that there was a period of almost exponential expansion (inflation) in the early universe, lasting from some initial time $t_i$ to a later time $t_R$ when a transition to the radiation phase of Standard Big Bang (SBB) cosmology occurs.  A space-time sketch of inflationary cosmology is shown in Fig. 3.  The vertical axis is time, and the horizontal axis is the physical length coordinate. During the period of inflation the horizon expands exponentially while the Hubble radius remains constant. Hence, the first of the above criteria is satisfied. If the period of inflation is sufficiently long, then all scales up to the current Hubble radius originate inside the Hubble radius at the beginning of inflation, thus allowing for a causal generation mechanism of fluctuations.  As first discussed in \cite{Mukh} (and in \cite{Starob} for gravitational waves), it is quantum vacuum perturbations which lead to the fluctuations which are currently observed.  Since the energy density during inflation is approximately constant,  fluctuations generated at all times have approximately the same amplitude \cite{Press}. As shown in \cite{Mukh}, the mechanism produces a slight red tilt of the spectrum.

\begin{center}
\begin{figure}[!htb]
\includegraphics[width=6.4cm]{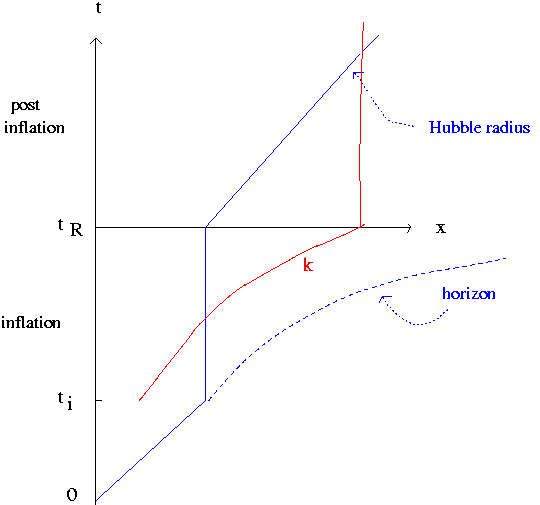}
\caption{Space-time sketch of an inflationary universe. The vertical axis is time, with the interval $t_i < t < t_R$ being the period of inflation. The horizontal axis indicates physical distance. Three different length scales are shown: the horizon (dashed curve which increased during inflation), the Hubble radius (constant during inflation) and the physical wavelength of a fluctuation mode labelled by $k$. If inflation lasts sufficiently long, all scales being observed today originate inside the Hubble radius at the beginning of inflation, and a causal generation mechanism for fluctuations is possible.}
  \label{fig3}
\end{figure}
\end{center}

Bouncing cosmologies (see \cite{bouncerev} for reviews) provide a second scenario in which all of the above criteria for a successful early universe scenario can be satisfied. Here, it is assumed that the universe starts in a contracting phase. Then, once the density has reached the string scale, some new physics (which cannot be described using an effective field theory approach in the narrow sense described in the following section) causes the transition to the expanding phase of SBB cosmology. The resulting space-time sketch is shown in Fig. 4. Here, the vertical axis is time, with $t = 0$ being the bounce time. The horizonal axis is comoving length (in terms of which the scale of the fluctuations is constant). Since time begins a $- \infty$, the horizon is infiinite and the first of the above criteria is trivially obeyed. It is obvious that all length scales originate inside the Hubble radius at sufficiently early times. Hence, the second criterion is also satisfied.  If one starts with quantum vacuum fluctuations on sub-Hubble scales in the far past, then there are classes of bouncing cosmologies in which a nearly scale-invariant spectrum is obtained. One example is the {\it matter bounce} scenario \cite{FB} in which the contracting phase is taken to be the mirror inverse of the SBB phase of expansion. A slight red tilt of the spectrum is obtained \cite{Ed} if the contribution of the observed dark energy component (treated as constant in time) is taken into account.  However, this scenario is unstable towards the development of anisotropies \cite{Peter}. The Ekpyrotic contraction scenario \cite{Ekp}, on the other hand, is a global attractor in initial condition space \cite{Erickson}. Depending on the matching conditions at the bounce (see e.g. \cite{DV} for a detailed discussion), a scale-invariant spectrum of curvature perturbations after the bounce can emerge if one begins with quantum vacuum fluctuations at the beginning of the contracting phase.  A recent example where a scale-invariant spectrum of curvature perturbations with a slight red tilt emerges is the recently proposed S-brane bounce model \cite{Ziwei}. 
 
 \begin{center}
\begin{figure}[!htb]
\includegraphics[width=6.4cm]{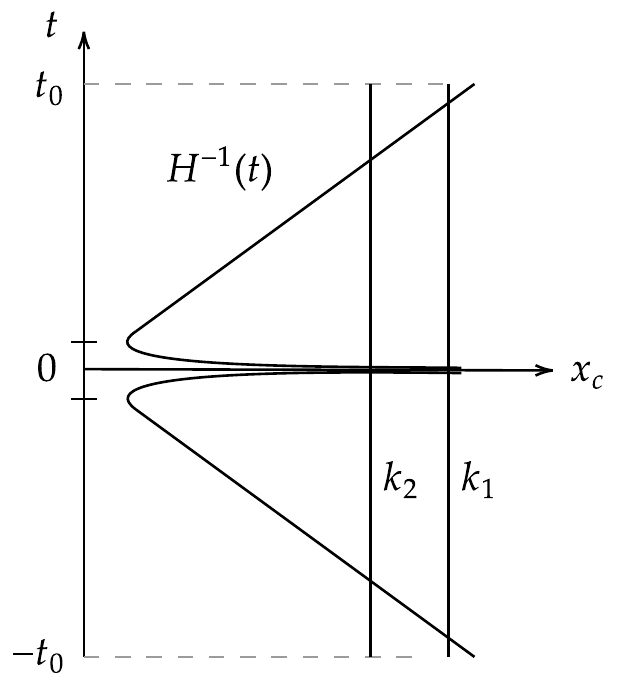}
\caption{Space-time sketch of a bouncing universe. The vertical axis is time, with $t = 0$ denoting the bounce point, the transition between contraction and expansion.  In this figure, the horizontal axis indicates comoving distance.  In comoving coordinates the wavelengths of fluctuation modes $k_1$ and $k_2$ are constant.  The space-time sketch corresponds to a bouncing scenario in which the contracting phase is the mirror inverse of the Standard Big Bang phase of expansion. The Hubble radius $H^{-1}(t)$ decreases linearly in time in the contracting phase while increasing during the phase of expansion. It is obvious that all scales being observed today originate inside the Hubble radius at some sufficiently early time in the past, and hence a causal generation mechanism for fluctuations is possible. Note that if the evolution of the cosmological scale factor at the bounce is analytic, then there is a time when $H = 0$ and the Hubble radius diverges.}
  \label{fig4}
\end{figure}
\end{center}

 The third scenario in which all three conditions required for a successful early universe cosmology can be satisfied is the {\it emergent scenario}, by which we mean a scenario the universe begins in an ``emergent'' phase which can be modelled at the level of a background space-time metric to be quasi-static.  The emergent phase ends with a phase transition to the usual SBB expanding cosmology. The resulting space-time diagram is sketched in Fig. 5. Here, the vertical axis is time, with the time $t_R$ denoting the time of the phase transition. Since time runs from $- \infty$ to $\infty$, the horizon is infinite.  Since the emergent phase is modelled as quasi-static, the Hubble radius also goes to infinity as we enter this phase. Thus, the first two criteria for a successful early universe cosmology are satisfied.  As will be shown later, if there are thermal fluctuations in the emergent phase with holographic scaling (which is natural from the point of view of superstring theory),  then a scale-invariant spectrum of curvature fluctuations and gravitational waves results \cite{NBV}.
 
 \begin{center}
\begin{figure}[!htb]
\includegraphics[width=6.4cm]{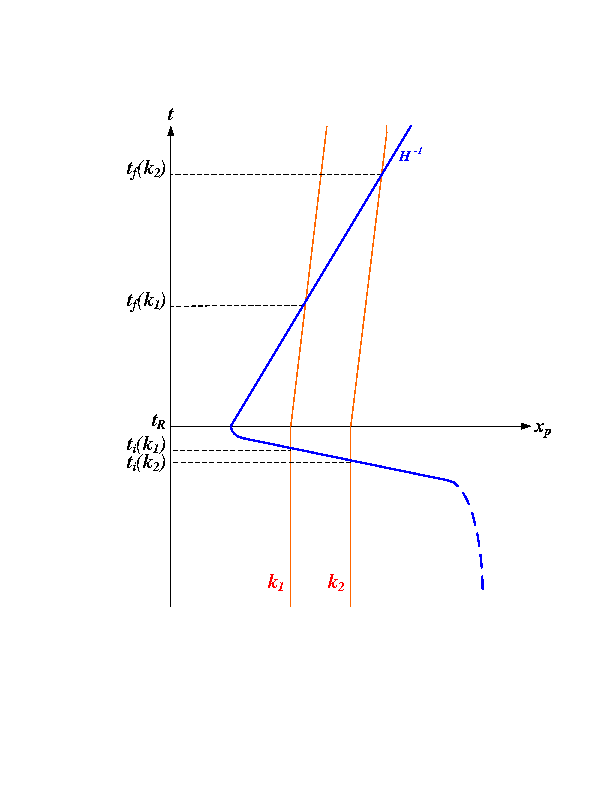}
\caption{Space-time sketch of an emergent scenario. The vertical axis is time, with the emergent phase being $t < t_R$. At the time $t_R$ as phase transition to the radiation phase of expansion of SBB cosmology occurs. The horizontal axis is physical distance. The solid thick (blue) curve labelled $H^{-1}$ is the Hubble radius, the thin (red) curves indicate the wavelength of two fluctuation modes, labelled $k_1$ and $k_2$, respectively. Since the Hubble radiusdiverges in the emergent phase, all scales originate inside the Hubble radius in the emergent phase and a casual generation mechanism of fluctuations is possible.}
  \label{fig5}
\end{figure}
\end{center}

 Bouncing and emergent universe scenarios cannot be obtained in the context of simple point particle-based effective field theories (EFT) - matter fields obeying the Null Energy Condition coupled to the Einstein-Hilbert action for space-time, since in this context it follows from (\ref{FRW1}) and (\ref{FRW2}) that it is impossible to obtain $H = 0$ and ${\dot H} > 0$.  Thus, to obtain realizations of bouncing and emergent scenarios it is necessary to go beyond an EFT analysis.  In the following section I will argue that we also need to go beyond an EFT analysis to describe cosmologies with a long phase of acceleration.
 
\section{Breakdown of Effective Field Theory}
\label{section3}

\subsection{Trans-Planckian Censorship Conjucture}

 Returning to the space-time sketch of inflationary cosmology (redrawn in Fig.  6), it is clear that if the period of inflation last for many e-foldings, then the wavelengths of all fluctuation modes which are currently probed in experiments are smaller than the Planck length at the beginning of the period of inflation. As pointed out in \cite{Jerome}, this implies that modes originate in a region in which quantum gravity effects must be taken into account.  From the point of view of an EFT treatment this can be called a {\it trans-Planckian problem}, but for physicists working on quantum gravity this might be viewed as a {\it trans-Planckian window of opportunity} to probe Planck-scale physics in observations (see \cite{JeromeRev} for a review of the large body of work devoted to this issue).
 
 \begin{center}
\begin{figure}[!htb]
\includegraphics[width=6.4cm]{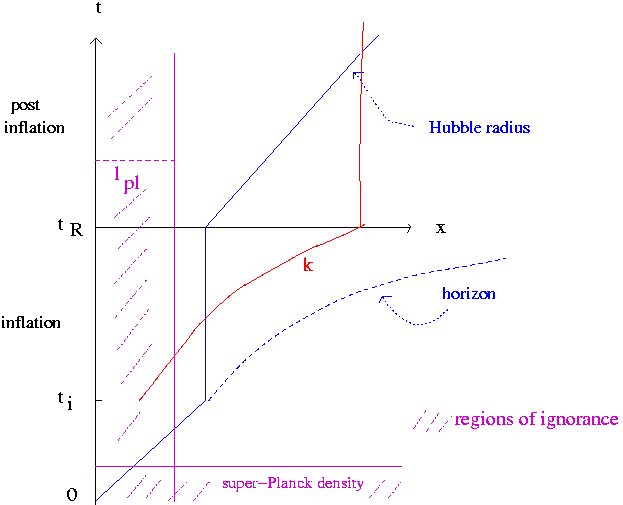}
\caption{Space-time sketch of an inflationary universe. The vertical axis is time, with the interval $t_i < t < t_R$ being the period of inflation. The horizontal axis indicates physical distance.  In addition to the length scales drawn in Fig.  \ref{fig3}, the Planck length is shown. If the period of inflation last sufficiently long, then the scale corresponding to the current Hubble radius (and all smaller scales) originate with a wavelength smaller than the Planck length at the beginning of inflation.}
  \label{fig6}
\end{figure}
\end{center}

Recently,  however, a conjecture was put forwards \cite{TCC} that the situation sketched in Fig. 6 cannot occur in any consistent theory of quantum gravity. Specifically, the conjecture states that no scale which initially was trans-Planckian can ever evolve to exit the Hubble horizon. This {\it Trans-Planckian Censorship Conjecture} (TCC) reads
 \be \label{TCC}
 \frac{a(t_R)}{a(t_i)} l_{pl} \, \leq \, H^{-1}(t_R) 
 \ee
 for any initial time $t_i$ and final time $t_R$. In Standard Big Bang cosmology the Hubble radius expands faster than fixed comoving scales, and hence the TCC imposes non constraints. As is obvious from the sketches of Figs. 4 and 5, the TCC also does not impose any restrictions provided that the energy scale of the bounce or the emergent phase is lower than the Planck scale. However, in the case of accelerating cosmologies the TCC implies an upper bound on the duration of the accelerating phase.  One immediate consequence is that Dark Energy cannot be a cosmological constant since in that case the period of accelerated expansion would be unbounded from above, and eventually scales which are currently trans-Planckian would exit the Hubble radius.
 
\begin{center}
\begin{figure}[!htb]
\includegraphics[width=6.4cm]{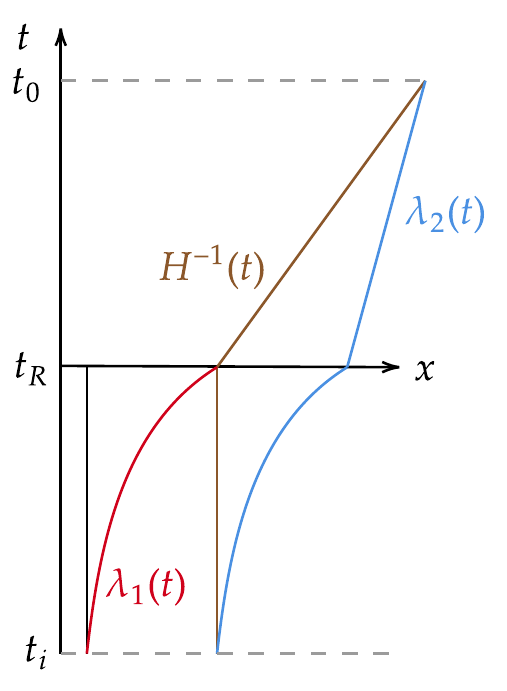}
\caption{Space-time sketch of an inflationary universe. The vertical axis is time, with the interval $t_i < t < t_R$ being the period of inflation. The horizontal axis indicates physical distance.   The curve $\lambda_1(t)$ indicates a fixed comoving scale whose physical wavelength equals the Planck length at $t_i$ which must remain inside the Hubble radius at all times and in particular at the end of the period of inflation, the curve $\lambda_2(t)$ is the physical wavelength corresponding to the current Hubble radius which must emerge from inside of the Hubble radius at the beginning of inflation.  In this sketch, the upper and lower bounds on the duration of inflation are shown to be marginally satisfied. This implies that the Hubble radius during inflation must be several orders of magnitude larger than what is usually assumed in single field inflation models.}
  \label{fig7}
\end{figure}
\end{center}

 For inflationary cosmology the TCC leads to severe constraints \cite{TCC2}.  The TCC condition (\ref{TCC}) (setting $t_i$ to be the beginning of the inflationary phase and $t_R$ its end time) imposes an upper bound on the duration of inflation, while there is a lower bound from demanding that the comoving scale corresponding to the current Hubble radius $H_0^{-1}$ originates iinside the Hubble radius at the beginning of inflation (at time $t_i$), i.e.
 \be \label{TCC2}
 \frac{a(t_i)}{a(t_0)} H_0^{-1} \, < \, H^{-1}(t_i) \, .
 \ee
Whether the two bounds are compatible depends on the energy scale $\eta$of inflation.  In the case of almost exponential inflation the compatibility of (\ref{TCC}) and (\ref{TCC2}) yields the constraint
\be
\eta \, < 3 \times 10^9 {\rm{GeV}} \, ,
\ee
a value several orders of magnitude lower than what is required in simple single field models of inflation. For such a low scale of $\eta$,  it takes tuning of the inflationary model to obtain a sufficiently large amplitude of cosmological fluctuations. If one succeeds at achieving this, the induced primordial gravitational waves would be negligible, with a tensor to scalar ratio $r$ of $r \sim 10^{-30}$. The bounds are slightly relaxed for power law inflation \cite{relaxed}, while they are strengthened if one takes into account a radiation phase before the onset of inflation \cite{strong}. This problem is illustrated in Fig. \ref{fig7} which shows the space-time sketch of inflation, the evolution of the mode with a wavelength equal to the Planck length at the beginning of inflation, and the wavelength of the mode whose current wavelength equals the current Hubble radius. In this sketch, the upper and lower bounds on the duration of inflation are shown to be both marginally satisfied.  As the calculation shows, this implies that the Hubble radius during the inflationary phase has to be several orders of magnitude larger than what is required in canonical single field models of inflation.

There are a number of justifications for the TCC which are independent of superstring theory \cite{TCC3}. The first is an analogy with Penrose's black hole Cosmic Censorship Hypothesis \cite{CCH} (CCH) which states that timelile singularities must be hidden from a distant observer by a horizon. For example, although Einstein's equations as EFT equations admit black hole solutions with charge $Q$ greater than the mass $M$, solutions which have naked singularities,  the CCH implies that in a consistent theory which completes the EFT of General Relativity such solutions cannot occur. We can translate this conjecture to cosmology: position space for black hole physics becomes momentum space for cosmology, the role of the black hole singularity is played by the set of trans-Planckian modes,  and the black hole horizon is replaced by the Hubble horizon.  Under this translation scheme, Penrose's CCH becomes precisely the TCC. 

The second justification is a unitarity argument. In an EFT approach we expand all fields in {\bf comoving} Fourier modes $k$ and quantize each mode as a harmonic oscillator. To avoid the UV singularity a UV cutoff is required, and this cutoff must be fixed in physical coordinates. In an expanding universe, this implies continuous creation of comoving modes, a clear violation of unitarity \cite{Weiss} (see also \cite{Cotler}).  Recall that the canonically normalized fluctuation variables oscillate on sub-Hubble scales but become squeezed and classicalize on super-Hubble scales. A conservative requirement is that the modes which can classicalize are insensitive to the non-unitarity of the EFT description. This is precisely the TCC condition (\ref{TCC}).

The third argument is a thermodynamic one. In a similar way in which the entanglement between modes inside and outside a black hole horizon leads to the breakdown of the semiclassical analysis of Hawking radiation after the {\it scrambling time} (see e.g. \cite{BHentanglement}),  the entanglement entropy density between sub- and super-Hubble modes (first discussed in \cite{Tom}) builds up during a period of accelerated expansion since the phase space of super-Hubble modes in increasing. Demanding that the entanglement entropy remains smaller than the thermal entropy density one would like to have after inflation leads to an upper bound on the period of inflation which is consistent with the TCC \cite{Brahma}.

The TCC also has implications for late time cosmology \cite{VS}: it implies that Dark Energy cannot be a bare cosmological constant, and it also imposes constraints on quintessence models of Dark Energy \cite{Lavinia}.

There is an interesting connection between the breakdown of EFT in cosmology and the quantum aspect of the cosmological constant problem. In an EFT approach, all fields are expanded in Fourier modes, each Fourier mode is quantized as a harmonic oscillator and thus contains a ground state energy. Adding up all of these ground state energies yields the famous cosmological constant problem \cite{CCproblem}. If, on the other hand, the correct approach to a quantum theory of space-time does not involve quantiziing harmonic oscillators, then the cosmological constant problem may not appear. An example where this is realized is the matrix model cosmology approach described in Section \ref{section6}.

\subsection{Swampland Constraints}

At the level of point particle EFTthere is an immense swampland for models.  There is no criterion for the number of space-time dimensions, no limit to the number of fields, no constraints on the range of these fields, and no constraints on the potential energy functions. Superstring theory imposes tight constraints on the set of EFTs which are consistent of string theory. These are the {\it swampland constraints}. (see e.g. \cite{swamprevs} for reviews of the swampland program). The set of EFTs which are consistent with string theory forms a small landscape of models in the much larger set of EFTs which one can write down. 

First of all, superstring theory fixes the number of space-time dimensions to be $10$. At first sight, this may be viewed as a serious problem for string theory.  To obtain a viable cosmology from superstring theory one must, therefore, find a mechanism to keep six of the nine spatial dimensions small. This is one aspect of the {\it moduli stabilization} issue reviewed at length in \cite{Review}.  As proposed initially in \cite{BV}, if one starts at string densities with all spatial dimensions being compact, then there is a natural mechanism which allows precisely three of the spatial dimensions to become large, while keeping the others confined to the string scale. This argument will be reviewed in the following section of this article.

In an EFT derived from superstring theory, all scalar fields have a stringy origin. Examples include the canonical field associated with the radius of an extra dimension, or with the position of a brane.  As was conjectured in \cite{distance} based on a set of representative examples from string theory, the field range $\delta \varphi$ of any canonically normalized scalar field $\varphi$ arising in a low energy EFT is bounded from above by 
\be \label{dist}
| \delta \varphi | \, < \, {\cal{O}}(1) m_{pl} \, .
\ee
This condition immediately puts pressure on models of cosmological large field inflation, the set of inflationary modes which do not require initial condition fine tuning (see e.g. \cite{ICrev} for a review of the initial condition problem for inflation). The origin of (\ref{dist}) comes from the fact that as $\varphi$ changes, the geometry of the string compactification changes and new stringy degrees of freedom can become effectively massless and need to be included in the low energy effective description. 

The potentials for scalar fields in an EFT coming from string theory are not arbitrary. They are induced by string processes, e.g. moduli stabilization mechanisms. Based on examples from  string theory, the {\it de Sitter conjecture} \cite{dS1} states that the slope of any potential $V(\varphi)$ which dominates the energy content of the universe must be either sufficiently steep
\be \label{cond1}
\vert \frac{m_{pl} V^{\prime}}{V} \vert \, > \, {\cal{O}}(1) \,  
\ee
or \cite{Krishnan, dS2} sufficiently tachyonic
\be \label{cond2}
\vert \frac{m_{pl}^2 V^{\prime \prime}}{V} \vert \,  > \, {\cal{O}}(1) \, ,
\ee
where a prime denotes a derivative with respect to $\varphi$. These conditions can be derived from entropy arguments \cite{dS2}, and they generalize to negative potentials \cite{HB}.

The swampland constraints  indicate that standard single field slow-roll inflation is difficult to obtain from string theory if one sticks to an EFT analysis. The swampland constraints can be relaxed in one makes use of {\it warm inflation} \cite{warm} or {\it gauge-flation} \cite{chromo}. In both these cases there are extra terms in the equation of motion for the inflaton field (the field driving inflation) which allows an inflationary trajectory on a potential which obeys (\ref{cond1}).

There has been a lot of work on inflationary model building in the context of effective field theories motivated by string theory (see e.g. \cite{Baumann} for a review),   but the consistency of these models at the level of full string theory is in doubt (see e.g. \cite{Tatar}),  and they are also inconsistent with the TCC coonstraint.

There are interesting attempts to obtain a finite duration phase of inflation based on coherent states in string theory \cite{Keshav}. The resulting scenario is consistent with the upper bound from the TCC on the duration of inflation.  A construction of a metastable de Sitter phase based on a coherent state of gravitons is given in \cite{Dvali}.

On the other hand, it is interesting to explore the possibility that string theory might lead to an early universe scenario which does not involve a period of cosmological inflation.  

\section{String Gas Cosmology} 
\label{section4}
 
 \subsection{String Gas Cosmology Background}
 
 String Gas Cosmology \cite{BV} (see also \cite{KP} for initial ideas, and \cite{SGCrevs} for reviews) is a model of early universe cosmology which makes use of the new degrees of freedom and new symmetries which differentiate string theory from point particle field theories.  We will consider a background space which for simplicity we take to be $T^9$ (a nine dimensional torus) with radius $R$ (in each dimension), and a gas of closed strings which lives in this space.
 
 Closed strings have the usual momentum mode degrees of freedom (like point particles) whose energies are quantized in units of $1/R$, i.e. (in string units)
 \be
 E_n \, = \, \frac{n}{R} \, ,
 \ee
 where $n$ is an integer. But there is also an infinite tower of string oscillatory modes (whose energies are independent of $R$), and string winding modes whose energies are quantized in units of $R$, i.e.
 \be
 E_m \, = \, m R \, ,
 \ee
 where $m$ is another integer. We see immediately that there is a new symmetry: replacing $R$ by $1/R$(in string units), and interchanging $n$ and $m$,  the spectrum of string states is unchanged. This symmetry extends to string interactions, and is believed to be fundamental also beyond perturbative string theory.  The duality symmetry implies that physics on small tori is equivalent to physics on large tori.  It is part of the larger T-duality symmetry group of string theory \cite{Tdual}.
 
\begin{center}
\begin{figure}[!htb]
\includegraphics[width=9cm]{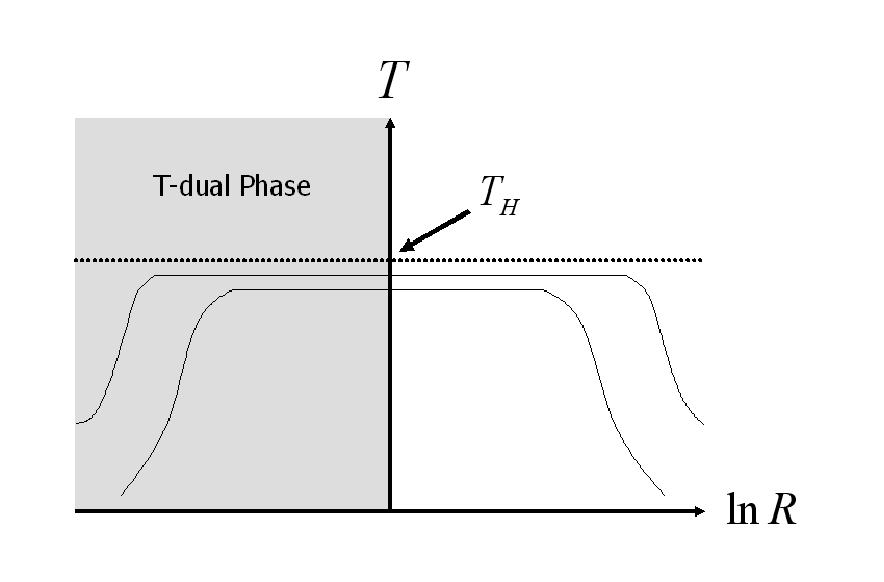}
\caption{Temperature (vertical axis) of a gas of closed strings in thermal equilibrium as a function of the radius of the box (horizontal axis). The length of the flat region depends on the entropy of the system. }
  \label{figSGC}
\end{figure}
\end{center}

 Considering a box of closed strings in thermal equilibrium, it has been known  for a long time \cite{Hagedorn} that there is a maximal temperature, the {\it Hagedorn temperature} $T_H$.  If $R \gg 1$, the energy of string gas will be in the momentum modes which are light. As the box shrinks and the energy density rises, then initially the temperature $T$ of the gas will rise as the energy of each string increases. Once the density approaches the string density (which will happen when the temperature is close to $T_H$, it becomes possible to excite oscillatory modes.  The energy of the system will drift into oscillatory modes while the temperature $T$ of the system remains almost constant. Once $R$ drops below $R = 1$, the energy will drift into the winding modes which are light for $R < 1$. The resulting temperature curve obeys the symmetry (see Figure \ref{figSGC})
 \be
 T(R) \, = \, T(1/R) \, .
 \ee
The size of the flat region of the curve depends on the total entropy of the gas - the larger the entropy, the larger the size. The flat region is called the {\it Hagedorn phase}.
 
 At the self-dual point $R = 1$, space must remain static. Based on this fact it was argued in \cite{BV} that even if we move away from the symmetry point but remain in the Hagedorn phase space-time will be quasi-static.  Energy considerations suggest that winding modes will prevent expansion while momentum modes prevent contraction. Hence, to obtain any large dimensions there needs to be a mechanism for winding modes about that dimension to annihilate.  In the absence of a chemical potential there will be an equal number of winding and anti-winding modes. If they intersect they can annihilate into string loops.  Superstrings are relativistic and trace out two-dimensional world sheets.  In the absence of long range forces between strings, there is vanishing interaction probability of the string world sheets in more than three spatial dimensions. Hence, as discussed in \cite{BV},  String Gas Cosmology (SGC) predicts that out of the nine dimensions of space,  only three can become larger. The annihilation of the winding modes allows a three-dimensional subspace to make the transition from a quasi-static Hagedorn phase to the radiation phase of Standard Big Bang expansion, while the other six dimensions of space remain confined at the string scale. This $SO(9) \rightarrow SO(6) \times SO(3)$ symmetry breaking will appear again in matrix cosmology.   
 
 The dynamics of the four-dimensional space-time emergiing from SGC leads to precisely the space-time diagram sketched in Figure. \ref{fig5}. SGC is hence a prototypical example of an emergent cosmological scenario.  As already indicated above, the presence of both momentum and winding modes about the compact dimensions leads to the stabilization of the size \cite{size} and shape \cite{shape} of the six-dimensional internal manifold. At late times, the effect of the string modes on the scalar field describing the size the internal dimensions (the ``radion'') is described by a potential which has a minimum at a value of the radion corresponding to the string length \cite{size2}. In the case of Heterotic string theory it can be shown\cite{size2} that the value of the potential at its minimum vanishes, and that its curvature is consistent with observational constraints. 
 
 At the level of effective field theory, it is a major challenge to understand the stabilization of string theory moduli fields (see e.g. \cite{Review} for an in-depth review). As argued above,  if we keep the truly string degrees of freedom, it is straightforward to see that the size and shape moduli of the extra dimensional manifold can be stabilized.  In order to produce a successful phenomenology, the string theoretic dilaton must also be stabilized. In the context of SGC this is more challenging. But it can be shown \cite{Frey} that the non-perturbative mechanism of gaugino stabilization \cite{gaugino} can be employed to stabilize the dilaton \footnote{This is the same mechanism which is invoked in many EFT approaches to string cosmology to stabilize the size moduli fields.}, without destabilizing the size and shape moduli.  On the other hand, gaugino condensation leads to \cite{Wei} supersymmetry breaking, at a scale which is typically of the string scale.


 \subsection{Fluctuations from String Gas Cosmology}

In inflationary cosmology it is assumed that fluctuations start out in their vacuum state \cite{Mukh, Starob}. This is justified if the period of inflation lasts a long time and hence redshifts all inhomogeneities pre-existing at the beginning of inflation. In SGC, howver, the initial state is a thermal state and there is no period of inflation. Hence, the dominant fluctuations will be thermal.

The energy density and pressure fluctuations of thermal system can be computed starting from the finite temperature partition function of the string gas. If we make the reasonable assumption that for the infrared modes (which are relevant for current cosmological observations) the equations for linearized Einstein gravity can be used to connect matter fluctuations to metric perturbations, then the curvature fluctuations are determined by the energy density perturbations, and the gravitational waves by the off-diagonal pressure perturbadtions.

Specifically, if we work in a coordinate system in which the scalar metric perturbations are diagonal, the {\it longitudinal gauge} (see \cite{MFB, RHBfluctsrev} for reviews), the metric including linear inhomogeneities takes the form
\begin{equation}
ds^2 \, = \, a^2(\eta) \bigl( (1 + 2 \Phi)d\eta^2 - [(1 - 
2 \Phi)\delta_{ij} + h_{ij}]d x^i d x^j\bigr) \,. 
\end{equation}
where $\Phi$ is the relativistic gravitational potential  and $h_{ij}$ is the transverse and traveless gravitational wave tensor which both depend on space and time \footnote{We are neglecting vector perturbations since they decay in an expanding universe.}. Then
\begin{equation}  \label{cosmopert}
\langle|\Phi(k)|^2\rangle \, = \, 16 \pi^2 G^2 
k^{-4} \langle\delta T^0{}_0(k) \delta T^0{}_0(k)\rangle \, , 
\end{equation}
and
\begin{equation} 
\label{tensorexp} \langle|h(k)|^2\rangle \, = \, 16 \pi^2 G^2 
k^{-4} \langle\delta T^i{}_j(k) \delta T^i{}_j(k)\rangle \, \,\, (i \neq j) \, .
\end{equation}

In a thermal system, the energy density fluctuations in a volume of radius $R = k^{-1}$ are given by the specific heat capacity $C_V$ via
\begin{equation} \label{spcheat}
\langle \delta\rho^2(k)  \rangle \,  = \,  \frac{T(k)^2}{R^6} C_V \, ,
\end{equation}
where $T(k)$ is the temperature when the scale $k$ crosses the Hubble radius.  For a thermal gas of strings on a compact space of radius $R$, the specific heat capacity has holographic scaling \cite{Tan}
\begin{equation}  \label{holo}
C_V(R(k))  \, \approx \, 2 \frac{R^2/\ell_s^3}{T(k) \left(1 - T(k)/T_H\right)}\, ,
\end{equation} 
where $l_s$ is the string length. To obtain this scaling, it is important to take into account that the string gas contains winding modes.  Since for subvolumes of a compact space there are also string modes which cross the space, we use the same formulae for subvolumes of the entire space \cite{NBV}.

Inserting (\ref{holo}) and (\ref{spcheat}) into (\ref{cosmopert}) yields the following result for the power spectrum of cosmological perturbations
\begin{eqnarray}
P_{\Phi}(k) \, 
&=& \, 8 G^2 k^{-1} <|\delta \rho(k)|^2> \nonumber \\ 
&=& \, 8 G^2 k^2 <(\delta M)^2>_R  \\ 
&=& \, 8 G^2 k^{-4} <(\delta \rho)^2>_R \nonumber \\ 
&=& \, 8 G^2 {T \over {\ell_s^3}} {1 \over {1 - T(k)/T_H}} \, .\nonumber 
\end{eqnarray}
Since the temperature in the Hagedorn phase is almost constant, $T(k)$ changes very little and the spectrum of cosmological perturbations is roughly scale-invariant. Looking a bit more closely, and referring back to figures \ref{fig5} and \ref{figSGC}, we see that smaller scales (corresponding to larger values of $k$) cross the Hubble radius at slightly later times and hence at slightly lower temperatures.  Thus, the resulting spectrum has a slight red tilt, like in the case of simple slow-roll inflationary models.  With the running of the spectrum, the predictions of SGC can be potentially distinguished from those of canonical inflation models \cite{SGCrunning}. There is, however, a much cleaner way to distinguish between the predictions of SGC and those of cosmological inflation - namely by considering the spectrum of the iinduced primordial gravitational waves \cite{BNPV}.

By taking the partial derivative of the string partition function with respect to $R$, we find the following relation for the diagonal pressure fluctuations
\begin{equation}
<|T_{ii}(R)|^2> \, \sim \, {T \over {l_s^3 R^4}} (1 - T/T_H) \, ,
\end{equation}
where, once again, $T = T(k)$. For thermal fluctuations we expect the off-diagonal pressure fluctuations to be of the same order of magnitude as the diagonal ones (see e.g. \cite{Nayeri}), and hence we obtain
\begin{eqnarray}
P_{h}(k) \, 
&=& \, 16 \pi^2 G^2 k^{-1} <|T_{ij}(k)|^2>  \nonumber  \\ 
&=& \, 16 \pi^2 G^2 k^{-4} <|T_{ij}(R)|^2>  \\ 
&\sim& \, 16 \pi^2 G^2 {T \over {\ell_s^3}} (1 - T/T_H) \nonumber 
\end{eqnarray}
for the power spectrum of gravitational waves.  To first approximation, this spectrum is scale-invariant. However, due to the fact that the factor of $(1 - T/T_H)$ is now in the numerator and not in the denominator, we find a slight blue tilt.  Since models of inflation based on General Relativity coupled to matter obeying the usual energy conditions always lead to a red tilt of the spectrum of gravitational waves, we have distilled a distinctive signature of SGC.  To leading order, SGC predicts the following consistency relation between the tilts of the scalar and tensor spectra \cite{BNP}
\be
n_t \, = \, 1 - n_s \, ,
\ee
where we recall that the scalar and tensor tilts $n_s$ and $n_t$ are defined via
\ba
P_{\Phi}(k) \, &\sim& \, k^{n_s - 1} \, \\
P_h(k) \, &\sim& \, k^{n_t} \nonumber \, .
\ea

Note that for the calcutations of the power spectra of cosmological perturbations and gravitational waves to work as indicated above, it is crucial that the dilaton is stabilized. If the dilaton were fluctuating itself, the results would be rather different \cite{object, reply}. However, as indicated at the end of the previous subsection, there is a natural dilaton stabilization mechanism which is operative in SGC.

The challenge for SGC is to provide a mathematical description of the Hagedorn phase.  What are the background equations that can lead to such a phase? It is clear that they are not those of General Relativity(GR),  but we should not expect the equations of GR to apply since they are inconsistent with the basic T-duality symmetry of string theory.  There has recently been an interesting proposal \cite{Vafa} of a new topological phase playing the role of the Hagedorn phase in the SGC toy model.  In Section \ref{section6} we will summarize a different approach \cite{BBL} based on the BFSS matrix model \cite{BFSS}.

\section{Pre-Big Bang Cosmology,  Ekpyrotic Cosmology and Double Field Theory}
\label{section5}
 
\subsection{Pre-Big-Bang Cosmology}

Pre-Big-Bang (PBB) cosmology \cite{PBB} (see \cite{PBB-Review} for a review) is a proposal to obtain an effective field theory description which is consistent with the basic T-duality symmetry of superstring theory.  The starting point of the PBB scenario is the fact that the massless sector of perturbative string theory contains, in addition to the graviton, $g_{\mu \nu}$ also a scalar field $\phi$ (the dilaton) and the antisymmetric tensor field (plus the corresponding fermionic fields).  

As a first approach to obtain a homogeneous and isotropic cosmology from the massless sector of string theory one can set the antisymmetric tensor field to zero. The action for the remaining bosonic fields is (working in four space-time dimensions)
\be \label{SFaction}
S(g_{\mu \nu}, \phi) \, = \, \int d^4x \sqrt{-g} e^{- \phi} 
\bigl[ R + g^{\mu \nu} \partial_{\mu} \phi \partial_{\nu} \phi  \bigr] \, ,
\ee
where $g$ is the determinant of the metric. and $R$ is the Ricci scalar of this metric  For comparison with observations, it is useful to work in the Einstein frame in which the action takes the form
\be \label{EFaction}
S(g^E_{\mu \nu},  \phi^E) \, = \, \int d^4x \sqrt{- g^E}
\big[ R^E- \frac{1}{2} \partial_{\mu} \phi^E\partial_{\nu} \phi^E \bigr] \, ,
\ee
where $g^E$ and $R^E$ are the determinant of the rescaled metric $g^E_{\mu \nu}$ and $R^E$ is its Ricci scalar. The variable transformation from the string frame to the Einstein frame is
\ba
g^E_{\mu \nu} \, = \,  e^{- \phi} g_{\mu \nu} \nonumber \\
\phi^E \, = \, \phi \, .
\ea

For a homogeneous and isotropic background we can work with the metric
\be
g_{\mu \nu} \, = \, {\rm{diag}} \bigl( 1,  -a^2(t)\delta{ij} \bigr) \, ,
\ee
where $t$ is time, and $a(t)$ is the string frame scale factor.  In terms of these variables, the T-duality symmetry of string theory is reflected in the {\it scale factor duality} symmetry \cite{SFduality}
\ba \label{duality}
a \, & \rightarrow &\, {\tilde{a}} \equiv a^{-1} \\
\phi \, & \rightarrow & \, {\tilde{\phi}} \equiv \phi - 3 {\rm{ln}}a \, . \nonumber
\ea
Under this duality, the rescaled dilaton
\be
{\bar{\phi}} \, \equiv \, \phi - 3 {\rm{ln}} a \,
\ee
is invariant.

It was observed \cite{PBB} that under the duality transformation (\ref{duality}) an expanding solution for $t > 0$ with fixed dilaton is mapped into a super-exponentially expanding solution for $t < 0$ with increasing dilaton.  However, without adding extra terms to the action, both solutions blow up at $t = 0$. In the Einstein frame, the solution for $t < 0$ corresponds to a contracting cosmology with an effective matter equation of state $w = 1$, where $w = p / \rho$, $w$ and $\rho$ being pressure and energy density, respectively. This equation arises since the only matter considered is the dilaton, and a homogeneous dilaton configuration only has kinetic energy.  Thus, in the Einstein frame the action of Pre-Big-Bang cosmology leads to a bouncing cosmology. The bounce, however, in a singular one in the absence of the addtion of extra terms to the action, terms which must violate the usual matter energy conditions from the point of view of the Einstein frame.

The action of Pre-Big-Bang cosmology is an effective action, and we expect this action to break down when the typical space-time curvature reaches the string scale.  String theory correction terms may well smooth out the singularity, similar to what happens in the Ekpyrotic scenario which will be discussed below.  Since the dilaton is increasing as the bounce point is approached, the string coupling constant increases. Thus, at the Pre-Big-Bang bounce, we expect both string corrections and coupling constant corrections to become crucial. In contrast, in the original Ekpyrotic scenario the string coupling constant tens to zero at the bounce point, and hence coupling constant correction terms are unimportant.

An important lesson which Pre-Big-Bang cosmology teaches us is that dynamical equations of motion for space-time which emergy from string theory may well lead to a cosmological scenario which does not involve a period of inflation.

The Pre-Big-Bang scenario faces several important challenges.  Firstly,  if we start out with cosmological fluctuations in their vacuum state (with a spectrum corresponding to $n_s = 2$), the fluctuations do not become scale-invariant on super-Hubble scales as the bounce point is approached.  Extra fields must be introduced to obtain a scale-invariant spectrum (see \cite{PBB-Review} for a review of various proposals).

The second challenge is that the energy density of the evolving dilaton condensate scales like that of anisotropies and spatial curvature. Hence,  unlike what happens in the Ekpyrotic scenario, spatial curvature and anisotropies are not diluted in the phase of contraction.  In order to obtain the low contribution of spatial curvature and anisotropies to the current expansion rate of the universe, a similar low contribution must be inserted at the corresponding time in the contracting phase. However, the situation is much better than in bounce scenarios with $w < 1$ (e.g. the matter bounce scenario of \cite{FB}) since in those scenarios anisotropies and curvature blow up during contraction \cite{Peter}.

\subsection{Ekpyrotic Scenario}

The Ekpyrotic scenario is a bouncing model involving a state of very slow contraction driven by (from the point of view of the Einstein frame) matter with an equation of state $w >> 1$. Such an equation of state can be obtained by invoking a scalar field with a negative (and fairly steep) exponential potential. of the form
\be \label{ekppot}
V(\phi) \, = \, - V_0 {\rm{exp}} (-  \sqrt{2/p} \phi / m_{pl} ) \,\,\,\,  p \ll 1 \, ,
\ee
where $V_0$ is a constant. Although the potential energy is negative, the total energy is positive in the scaling solution of the scalar field (the analog of the inflationary slow-roll trajectory) since the kinetic energy dominates over the potential. The solution for the cosmological scale factor is
\be
a(t) \, \sim \, (-t)^p 
\ee
with
\be
w \, \simeq \, \frac{4}{3p} \, \gg 1 \, .
\ee

The initial motivation for the Ekpyrotic scenario \cite{Ekp} came from heterotic M-theory \cite{HW},  an ansatz for a string vacuum in which one of the extra spatial directions is an orbifold which can be viewed as an interval with length greater than the string scale bounded on either side by boundary branes.  It was assumed that the scalar field is the canonical variable $\phi$ related to the time-dependent distance between a bulk brane and the boundary.  In another realization, $R$ was assumed to be the length of the interval, and this interval was undergoing cyclic evolution \cite{Ekp2}. The canonical variable $\phi$ is related to $R$ via 
\be
R \, \sim \, {\rm{log}}(\phi) \, .
\ee
A power law potential for $R$ then translates to an exponential potential for $\phi$. If the ground state of the system is AdS (anti-de Sitter) space,  then the total potential can be argued to be of the form (\ref{ekppot}).
 
 Since the potential is exponential, the energy density increases exponentially as $\phi$ decreases. Thus, the total field range during the relevant phase of cosmological contraction (between when cosmological and microphysical length scales cross the Hubble radius) is small compared to the Planck mass, and the scneario is consistent with the field range constraint. More specifically,   $\phi$ evolves as
\be
\phi(t) \, = \,\sqrt{2p} m_{pl} {\rm log} \bigl( - \sqrt{\frac{V_0}{m_{pl}^2 p (1 - 3p)}} t \bigr) 
\, .
\ee
Since $p \ll 1$,  the de Sitter criterion on the shape of the potential is obeyed (see \cite{HB} for the generatization of the de Sitter criterion to negative potentials).

The challenge for the Ekpyrotic scenario is,  like in Pre-Big-Bang cosmology, to obtain a nonsingular bounce. In addition, if one starts with cosmological perturbations in the far past in their quantum vacuum state, then the curvature fluctuations retain their vacuum spectrum on super-Hubble scales until shortly before the bounce. \cite{Ekp-flucts}.  Thus,  in order to become a viable alternative to inflation, there needs to be a mechanism to obtain a cosmological bounce, and to ensure that the primordial curvature fluctuations in the expanding phase are scale-invariant.  

One way to obtain scale-invariant curvature fluctuations  before the bounce is to invoke isocurvature fluctuations: one postulates the existence of a second scalar field with a negative exponential potential. Fluctuations of this field acquire a scale-invariant spectrum during Ekpyrotic contraction, and through a coupling of the two scalar fields scale-invariant curvature perturbations before the bounce can be induced \cite{twofield}.  There is, however, the danger that this coupling induces an instability of the Ekpyrotic field trajectory (see e.g. \cite{Ekp-Review} for a review).

The transfer of fluctuations from a contracting phase to an expanding phase is a tricky issue.  When discussing curvature fluctuations on length scales larger than the Hubble radius, one has to worry about gauge ambiguities. There are different variables which characterize curvature perturbations (see e.g. \cite{MFB}). One of the most common ones is the relativistic potential $\Phi({\bf{x}}, t)$ which is the metric perturbation in a gauge in which the metric, including scalar fluctuations, is diagonal
\be
ds^2 \, = \, (1 + 2\Phi) dt^2 - a(t)^2 (1 - 2\Phi) d{\bf x}^2 \, ,
\ee
where ${\bf{x}}$ are the comoving spatial coordinates. The second one is the variable $v({\bf{x}}, t)$ in terms of which the action for cosmological perturbations has canonical form.  In the case of inflationary cosmology, both of these variables have the same spectral slope. In the case of Ekpyrotic contraction, the spectrum of $\Phi$ becomes approximately scale-invariant \cite{Ekp3} while that of $v$ retains its vacuum shape. To agree with current observations,  we require the spectrum of $\Phi$ in the expanding phase to be scale-invariant (plus a slight red tilt).  Based on studies of toy models in which the background bounce is smoothed out by the addition of matter fields or modified gravity corrections (see e.g. \cite{bounce-toy}) it was believed that it is the variable $v$ which masses smoothly through the bounce from Ekpyrotic contraction to standard cosmology expansion.  If, however, the bounce is abrupt (continuous but non-differentiable evolution of the scale factor) and can be modelled by a space-like transition hypersurface, then, as discussed in \cite{DV}, which variable transits smoothly through the bounce depends crucially on the location of the boundary surface in space-time. In fact, for generic matching surfaces it is the variable $\Phi$ which transitions continuously (unlike what happens, for example, at the reheating surface in inflationary cosmology). 

The Ekpyrotic scenario is desccribed using effective field theory tools - a scalar field coupled to Einstein gravity. This formalism will break down once the energy density reaches the string scale at time $t_s$ at the end of the contracting phase. At that point,  new towers of string states become light and have to be included in the low energy effective action. A proposal for how to do this was put forwards in \cite{Ziwei}. The idea is that since the new terms becme important precisely at time $t_s$, they should be included in the action as a new term localized on the spatial hypersurface when the density equals the string density. Such a term is a field theory representation of a stringy S-brane \cite{Sbrane} (see also \cite{Kounnas} for previous uses of S-branes in string cosmology). It has vanishing energy density (since the component on the stress-energy tensor orthogonal to the surface vanishes) and negative pressure (as a relativistic object it has tension and hence negative pressure). Hence, an S-brane violates the usual matter energy conditions and can mediate the instantaneous transition between contraction and expansion \cite{Ziwei}, and, if the S-brane is located on a zero-shear surface, it can be shown that it is the variable $\Phi$ which passes smoothly through the bounce. Hence, a scale-invariant spectrum of curvature fluctuations with a slight red tilt results. The gravitational waves also acquire a roughly scale-invariant spectrum, but with a slight blue tilt \cite{Ziwei}.

Among bouncing cosmologies, those with a phase of very slow (Ekpyrotic) contraction are particularly promising since spatial curvature and anisotropies are diluted during the contracting phase.  The homogeneous Ekpyrotic trajectory is an attractor in initial condition space \cite{Erickson}.  In light of this, it would be of interest to devote more efforts to finding implementations of the scenario in string theory.

\subsection{Double Field Theory Cosmology}

Double field theory \cite{DFT} (see also \cite{DFT-Review} for a review) has a similar motivation to Pre-Big-Bang cosmology: one is searching for a low energy effective field theory for cosmology which is consistent with the key symmetries of string theory.  Instead of focusing only on the scale factor duality symmetry which underlies the Pre-Big-Bang scenario,  Double Field Theory (DFT) implements the full $O(D, D)$ T-duality symmetry of string theory., where $D$ is the number of space-time dimensions (in the Euclidean setting) or the number of spatial dimensions (in the Minkowski signature approach). 

DFT involves the doubling of the number of spatial dimensions. This can be motivated by stepping back and considering closed strings on a torus of radius $R$. As discussed in the section on String Gas Cosmology,  strings have both momentum and winding modes. In point particle quantum mechanics, the position eigenstate $\vert x >$ is defined through the momentum modes $\vert p >$:
\be
\vert x > \, = \, \sum_p e^{i p x} \vert p > \, .
\ee
For strings, we can define a dual position eigenstate $\vert {\tilde{x}} >$ defined in analogous fashion \cite{BV} through the winding modes $\vert w >$:
\be
\vert {\tilde{x}} > \, = \, \sum_w e^{i w {\tilde{x}}} \vert w > \, .
\ee
If $R$ is much larger than the string length, the light degrees of freedom are the momentum modes and hence it is the position eigenstate $\vert x >$ which is energetically accessible. On the other hand, if $R$ is much smaller than the string length, then it is the dual position eigenstate $\vert {\tilde{x}} >$ which is accessible.

DFT can be viewed as an effective field theory on the doubled space described above involving the fields which correspond to the massless modes of superstring theory, namely the dilaton $\phi$, the metric $g_{ij}$ and the antisymmetric tensor field $b_{ij}$. These fields are taken to be functions of the regular spatial coordinates ${\bf{x}}$ and the dual spatial coordinates ${\bf{\tilde{x}}}$.  The metric and the antisymmetric tensor field can be combined to form a ``generalized metric'' ${\cal{H}}_{MN}$ on the doubled space (the indices $M$ and $N$ run over the $2D$ indices of the doubled space
\ba
\mathcal{H}_{MN} &=& 
\begin{bmatrix}
g^{ij} & -g^{ik}b_{kj} \\
b_{ik}g^{kj} & g_{ij}-b_{ik}g^{kl}b_{lj} \\
\end{bmatrix} 
\ea
In a setup in which time is not dualized, then in terms of ${\cal{H}}$, the metric in a cosmological setting can be written as
\be
ds^2 \, = \, dt^2 - {\cal{H}}_{MN} dX^M dX^N \, ,
\ee
where the $X^M$ coordinates are the combined regular and dual spatial coordinates.
\be
X^M \, = \,  (\tilde{x}_i,x^{i}) \, .
\ee

The action of DFT is chosen such that it reduces - in the case when the fields do not depend on the dual spatial coordinates - to the usual supergravity action. The choice is
\be
S \, = \,  \int dxd\tilde{x} e^{- \Phi}\mathcal{R} \, ,
\ee
where $\Phi$ is the rescaled dilaton
\be
e^{-\Phi} \, = \, \sqrt{-g} e^{2\phi} \, ,
\ee
with $g$ being the determinant of the metric, and
\begin{eqnarray}
\mathcal{R} & = & \frac{1}{8}\mathcal{H}^{MN}\partial_{M}\mathcal{H}^{KL}\partial_{N}\mathcal{H}_{KL}-\frac{1}{2}\mathcal{H}^{MN}\partial_{M}\mathcal{H}^{KL}\partial_{K}\mathcal{H}_{NL}\nonumber \\
 &+&2\mathcal{H}^{MN}\partial_{M}\partial_{N}\Phi-\partial_{M}\partial_{N}\mathcal{H}^{MN}-\mathcal{H}^{MN}\partial_{M}\Phi\partial_{N}\Phi\nonumber \nonumber \\
 &+&2\partial_{M}\mathcal{H}^{MN}\partial_{N}\Phi  \\
 &+& \frac{1}{2}\eta^{MN}\eta^{KL}\partial_{M}\mathcal{E}_{\quad K}^{A}\partial_{N}\mathcal{E}_{\quad L}^{B}\mathcal{H}_{AB} \, .  \nonumber 
 \end{eqnarray}
Here, the tensor $\eta_{MN}$ is
\ba
\eta^{MN} &=&
\begin{bmatrix}
0 & \delta^{\;\;j}_i \\
\delta^{i}_{\;\;j} & 0 \\
\end{bmatrix} \, .
\ea

Considering the cosmological ansatz for the metric
\be
ds^2 \, = \, dt^2 - a(t)^2 d{\bf{x}}^2 - a(t)^{-2} d{\bf{\tilde{x}}}^2 
\ee
it can be shown that point particle geodesics can be continued to infinity in both time directions in a usual cosmological background where $a(0) = 0$. Key is the interpretation of the motion in terms of a dual time coordinate
\be
{\tilde{t}} \, = \, \frac{1}{t} 
\ee
for times $t \ll 1$, where here $t$ is a dimensionless time normalized such that density reaches the T-dual density \cite{Amanda1}. 

In another study \cite{Amanda2}, the equations of motion for dilaton gravity ($d = D - 1$ being the number of spatial dimensions)
\begin{eqnarray}
\left(\dot{\phi}-dH\right)^{2}-dH^{2} & = & e^{\phi}\rho \nonumber \\
\dot{H}-H\left(\dot{\phi}-dH\right) & = & \frac{1}{2}e^{\phi}p \nonumber \\
2\left(\ddot{\phi}-d\dot{H}\right)-\left(\dot{\phi}-dH\right)^{2}-dH^{2} & = & 0 \, 
\end{eqnarray}
were coupled to a matter source with an equation of state which reflects the T-duality symmetry:
\begin{equation}
w\left(a\right) \, = \, \frac{2}{\pi d}\arctan\left(\epsilon {\rm ln} \left( \frac{a}{a_0} \right) \right) \, ,
\end{equation}
where $\epsilon$ is a constant.This equation of state interpolates between winding domination for small $a$ to radiation domination for large $a$.

With the ansatz
\ba
a(t) \, & \sim & \, \bigl( \frac{t}{t_0} \bigr)^{\alpha}  \\
{\bar \phi}(t) \, & \sim & \, \beta\ln (t/t_0) \, , \nonumber
\ea
limiting solutions were found. In the Hagedorn phase when $w = 0$ the solutions had the power law indices $( \alpha, \beta ) \, = \, ( 0, 2)$ which corresponds to a static metric in the string frame, but with evolving dilaton. In the large $a$ phase with the radiation equation of state $w = 1/d$ the solutions are $( \alpha, \beta ) \, = \, \bigl( \frac{2}{D}, \frac{2}{D} (D - 1) \bigr)$ which corresponds to a radiation dominated universe with fixed dilaton. In the small $a$ phase with $w = - 1/d$ the solutions have $( \alpha, \beta ) \, = \, \bigl( - \frac{2}{D}, \frac{2}{D} (D - 1) \bigr) $.  In the string frame, these solutions correspond to nonsingular bouncing models.  This analysis shows that the fact that $a \rightarrow 0$ as $t \rightarrow 0$ does not correspond to a physical singularity.  This reinforces the lessons learned from String Gas Cosmology that including the T-duality symmetry into cosmology leads to a resolution of the Big Bang singularity of pure Einstein gravity.

Another approach to DFT is to include a dual time ${\tilde{t}}$ at the outset \cite{Amanda3}.  The cosmological metric is then
\be
ds^2 \, = \, dt^2 + d{\tilde{t}}^2 - a^2(t, {\tilde{t}})^2 dx^2 - a(t, {\tilde{t}})^{-2} d{\tilde{x}}^2 \, ,
\ee
where the scale factor now depends on both time variables.  Inserting this ansatz for the metric into the DFT action results in a set of Friedmann-like equations which involve both $H = {\dot{a}}/a$ and ${\tilde{H}} = a^{\prime}/a$, where an overdot represents the derivative with respect to time $t$, and a prime the derivative with respect to dual time.  We can once again couple the dualized geometry with matter which has an equation of state which reflects the T-duality symmetry, i.e. which is dominated by the momentum modes for $a \gg 1$ and by winding modes for $a \ll 1$. To extract physical meaning from these equations, it is important to impose the {\it{section condition}} (see e.g. \cite{DFT-Review}). In \cite{Amanda3} it was proposed to do this independently for $a \gg 1$ and $a \ll 1$. In the former case we assume that the variables do not depend on ${\tilde{t}}$, and in the latter case we assume that variables do not depend on $t$. The resulting cosmology is nonsingular \footnote{See also \cite{Park} for other approaches to DFT cosmology.}.

However,  without including string loop and string length ($\alpha^{\prime}$) \footnote{The string length $l_s$ is determined in terms of $\alpha^{\prime}$ via $l_s = \sqrt{2 \pi \alpha^{\prime}}$.} corrections, it was not possible to obtain a quasi-static Hagedorn phase (in which both the metric and the dilaton loiter). An obvious question is whether higher order corrections can change this conclusion. In a recent advance, it was shown how to include all order $\alpha^{\prime}$ corrections into the minisuperspace version (homogeneous and isotropic metric) of the DFT action \cite{Hohm} (see also \cite{Carmen} for some earlier studied of $\alpha^{\prime}$ corrections in DFT), and the analysis was generalized to include matter in \cite{HG1}. The starting point was the action for the massless degrees of freedom of string theory written in the form
\be
S_0 \, = \, \int d^dx I_0 \, ,
\ee
(where $n$ is the lapse function in the metric) with
\be
I_0 \, = \, - \frac{1}{2 \kappa^2} \int dt n e^{-\Phi} 
\bigl[ ({\cal{D}}\Phi)^2  + \frac{1}{8} tr ({\cal{D}} {\cal{S}})^2 \bigr]
\ee
where $\kappa$ is related to Newton's gravitational constant in the standard way, and the matrix ${\cal{S}}$ is
\be
{\cal{S}} \, = \, \eta {\cal{H}} 
\ee
in terms of the matices $\eta$ and ${\cal{H}}$ defined above. In \cite{Hohm} it was shown based on symmetry arguments that the most general cosmological action (i.e.  minisuperspace action for a FLRW cosmology) including all $\alpha^{\prime}$ corrections can be written in the form
\be
I \, = \,  - \frac{1}{2 \kappa^2} \int dt n e^{- \Phi} 
\bigl(  ({\cal{D}}\Phi)^2  + \sum_{k = 1}^{\infty} (\alpha^{\prime})^{(k - 1)} c_k tr ({\cal{D}} {\cal{S}})^{2k}  \bigr) \, ,
\ee
where the $c_k$ are coefficients which can only be determined by full string theory \footnote{See e.g \cite{Codina} for a study of this approach.} (but we know that $c_1 = 1/8$).  The resulting equations for a homogeneous and isotropic cosmology were studied in \cite{Hohm}, and they can be written in terms of a function
\be
f(H) \, = \, d \sum_{k = 1}^{\infty} (- \alpha^{\prime})^{(k - 1)} 2^{2(k + 1)} k c_k H^{2k -1}
\ee
which depends on the coefficients $c_k$. It was shown that this function $f(H)$ can be chosen such that de Sitter solutions exist. However, it is not clear whether the requirements on $f(H)$ to obtain such solutions are consistent with string theory, or whether they lie in the ``swampland''.

In \cite{HG1, HB-GF} solutions of the equations of motion of DFT cosmology from \cite{Hohm} coupled to matter obeying the T-duality symmetry were analyzed., and their stability was studied \cite{HB-GF} \footnote{See also \cite{WuYang} for cosmological solutions of the pure gravity-dilaton system.}. In \cite{HG2} (see also \cite{Meissner}) it was shown that the formalism of \cite{Hohm} (including matter) can be generalized to an anisotropic setting. In particular, we can consider a setup in which matter is initially dominated by winding modes about all spatial dimensions and then undergoes a phase transition in which the winding modes about three (external) dimensions decay into radiation while those about the other six (internal) dimensions persist, as argued to be the case in String Gas Cosmology. It was shown \cite{HG2} (see also \cite{Quintin}) that an interesting emergent scenario results in which the Einstein frame radius of the internal dimensions in constant both before and after the phase transition (with a slight change in size at the transition point) while the Einstein frame radius of the external dimensions is constant before the phase transition but grows as in a radiation-dominated universe after the transition. In the string frame, however,  there is a de Sitter phase of both the external and internal dimensions before the phase transition. After the phase transition the dilaton is stabilized, and hence there is no difference between the string and the Einstein frame.  Once again, the time evolution before the phase transition is not the same as what is postulated in String Gas Cosmology. However, the evolution after the phase transition is the same. In particular, it is confirmed that the radion stabilization mechanism \cite{size, size2} is operative. Note that the size and horizon problems are absent in the classical string geometry approach \cite{HG3}. In these studies, the antisymmetric tensor field was set to zero. For a study of the effects of including a non-vanishing $b_{ij}$ field see \cite{Paul}.

The formalism of \cite{Hohm} to include all $O(D, D)$ invariant $\alpha^{\prime}$ correction terms in the cosmological action has also recently been applied in \cite{GV} to construct non-singular dilaton gravity models which interpolate between an Einstein frame contacting phase and an Einstein frame phase of expansion and thus to provide a realization of a smooth PBB cosmology. 

\section{Nonperturbative Approaches}
\label{section6}

\subsection{Matrix Cosmology}

If superstring theory is to provide a consistent unified quantum desription of all four forces of nature, it must also provide a picture of space and time which is consistent with quantum mechanics. This (and also the problems discussed in Section III) requires an analysis which goes beyond perturbation theory.  Matrix theory and specifically the BFSS matrix model \cite{BFSS} (see \cite{Hoppe} for earlier related work, and \cite{other-mm} for some other matrix models related to string theory) is a proposed non-perturbative definition of superstring theory. It is a supersymmetric model involving $9$ bosonic matrices $X_i$ (called ``spatial'' matrices below) and $16$ fermionic $N \times N$ matrices.  The bosonic part of the BFSS Lagrangian is given by
\begin{equation}
L \, = \, \frac{1}{2 g^2} \bigl[ {\rm Tr} \bigl( \frac{1}{2} (D_\tau X_i)^2 - \frac{1}{4} [X_i, X_j]^2 \bigr) \bigr] \, ,
\end{equation}
where $i, j$ run from $1$ to $9$, and $D_\tau$ stands for the covariant derivative
\be
D_\tau X_i \, = \partial_\tau X_i - i [A_0(\tau), X_i] \, ,
\ee
with the square brackets indicating matrix commutators. In the above, a $10'th$ $N \times N$ matrix $A_0$ has been introduced (the ``temporal'' matrix).  The matrix model time is denoted by $\tau$, and indices are contracted with the usual $\eta_{\mu \nu}$ symbol. The Lagrangian contains the coupling constant $g$ (which is related to the string length $l_s$). The Lagrangian determines the BFSS action $S_{BFSS}$ in the usual way.

Based on perturbative calculations, there are detailed reasons to expect that it is precisely the specific matrix model given above which in the $N \rightarrow \infty$ limit (with the `t Hooft coupling $g^2 N$ held fixed) yields a non-perturbative definition of superstring theory (see e.g. \cite{Taylor,  Zarembo, Ydri} for reviews of the BFSS matrix model).

We \cite{BBL} will consider the matrix model in a high temperature state. In this case, we can expand the spatial matrices in Matsubara modes:
\begin{equation}
X_i(t) \, = \, \sum_n X_i^n e^{2 \pi i T t} \, .
\end{equation}
Inserting this expansion into the  bosonic action, we find that at high temperatures the action is dominated by the $n=0$ Matsubara modes:
\be \label{highT}
S^b_{BFSS} \, = \, S^b_{IKKT} + {\cal{O}}(1/T) \, ,
\ee
where $S^b_{IKKT}$ is the contribution of the $n = 0$ modes to the action. The superscript $b$ indicates that we are focusing on the bosonic part. 

As was realized in \cite{Kawahara}, under the change of variables
\begin{equation}
A_i \, \equiv \, T^{-1/4} X_i^0  
\end{equation}
the contribution of the $n = 0$ modes to the BFSS action is the same as the bosonic part of another matrix model action, the IKKT action \cite{IKKT}. The IKKT action 
\begin{equation} \label{IKKTaction}
S_{IKKT} \, = \,  -\frac{1}{g^2} {\rm{Tr}} \bigl( \frac{1}{4} [A^a, A^b] [A_a,A_b] + \frac{i}{2} \bar{\psi}_\alpha ({\cal{C}} \Gamma^a )_{\alpha\beta} [A_a,\psi_\beta] \bigr) \, ,
\end{equation}
where the index $a$ runs from $0$ to $9$, indices are again raised with the $\eta_{ab}$ symbol, the $\psi$ stands for a 16-component vector of Fermionic matrices,  ${\cal{C}}$ is the charge conjugation matrix, and the $\Gamma^a$ are the $10-D$ gamma-matrices, is a proposed non-perturbative definition of Type IIB superstring theory. The bosonic part of this action is the same as the $S^b_{IKKT}$ introduced in (\ref{highT}).

The IKKT action is a pure matrix model action defined via the partition function
\begin{equation}
Z \, = \, \int dA d\psi e^{iS} \, .
\end{equation}
In particular, there is neither time nor space.

The approach of \cite{BBL} (see \cite{BBL-Review} for short reviews) is to start with the BFSS model in a high temperature state and use the $n = 0$ modes of the matrices to extract an emergent space-time.  There is a prescription to obtain an induced spatial metric \cite{BBL2}.  In the $N \rightarrow \infty$ limit, the induced time is continuous and runs from $- \infty$ to $+ \infty$,  the induced space infinite and continuous in the same limit, and the induced spatial metric is flat.  Roughly speaking, the eigenvalues of the matrix $A_0$ yield the temporal elements and the eigenvalues of $A_i$ are related to the spatial dimensions. To be more specific, we work in the basis in which $A_0$ is diagonal (note that since the matrices $A_a$ are Hermitean, we can always choose such a basis).  It can be shown \cite{Ito} that
\be
\frac{1}{N} \bigl< {\rm Tr} A_0^2 \bigr> \, \sim \, \kappa N \, ,
\ee
where $\kappa \ll 1$ is a positive constant, and the brackets indicate the expectation value with respect to the matrix model measure.  From this it follows that the maximal eigenvalue tends to infinity 
\be
t_{max} \, \sim \, \sqrt{N} 
\ee
in the $N \rightarrow \infty$ limit, and the separation $\Delta t$ of eigenvalues tends to zero:
\be
\Delta t \, \sim \, \frac{1}{\sqrt{N}} \, .
\ee
Thus,  in the $N \rightarrow \infty$ limit, continous time which is both past and future infinite emerges.

Next, we turn to emergent space \cite{BBL2}.  Continuing to work in the basis in which $A_0$ is diagonal, we make use of the fact that the off-diagonal elements of the spatial matrices decay sufficiently far from the diagonal
\be \label{prop1}
\sum_i \bigl< |A_i|^2_{ab} \bigr> \, \rightarrow \, 0 \,\,\, {\rm{when}} \,\,\, |a - b| > n_c
\ee
 with
 \be
n_c \, \sim \, \sqrt{N} \, .
\ee
These two properties can be seen from the partition function using the Riemann-Lebesgue Lemma. Numerical simulations indicate that
\be \label{prop2}
\sum_i \bigl< |A_i|^2_{ab} \bigr> \,\,\, \sim \, {\rm{constant}} \,\,\ {\rm{when}} \,\,\, |a - b| < n_c \, .
\ee
The proposal of \cite{Nishimura} is to define spatial submatrices ${\bar{A}}_i(t, n_i)$ by taking the matrix $A_i$, running down the diagonal to a position which corresponds to the time $t$ of the $A_0$ matrix, and to take a $n_i \times n_i$ matrix centered at this point. Then,  following \cite{Nishimura, Nish}, one can define {\it extent of space} parameters
 \be
x_i(t, n_i)^2 \, \equiv \, \left\langle \frac{1}{n_i} \text{Tr} ({\bar{A_i}})(t, n_i))^2 \right\rangle \, ,.
\ee
As has been observed numerically, and confirmed using the Gaussian expansion method \cite{Lowe},  there is a phase transition in which the $SO(9)$ symmetry of the action is spontaneously broken to $SO(3) \times ``SO(6)''$ in the sense that at large values of $t$,  precisely three of the nine extent of space parameters become large while the other six remain small \cite{SBreaking} (quotation marks are used in the above to indicate that the internal space is some non-commutative space, not a smooth geometrical manifold.)This is reminiscent of the phase transition which happens in String Gas Cosmology in which the decay of string winding modes allows precisely three of the nine spatial dimensions to effectively decompactify. Note that the presence of fermions is crucial to have this phase transition (see e.g. \cite{Nish-Review} for reviews of the IKKT matrix model). 

Returning to the BFSS model, it can be shown, \cite{BBL3}  employing the same Gaussian expansion method  as used in the IKKT model,  that the $SO(9)$ symmetric state is not a minimum of the free energy. Based on the physical argument given in the previous paragraph we expect the state which minimizes the free energy to have $SO(3) \times ``SO(6)''$ symmetry, as in the case of the IKKT model. Note that fermions are again crucial to have symmetry breaking. There is no symmetry breaking in the pure bosonic version of the BFSS model \cite{BBL3}.

Focusing on the three-dimensional subspace of the large dimensions,  the proposal of \cite{BBL2} is to view $n_i$ as comoving spatial coordinates, and to define the length of a curve along the $i$ coordinate axis from the coordinate origin to coordinate $n_i$ via the extent of space parameter discussed above
\be
l_{phys, i}^2(n,_i  t) \, \equiv \, \left\langle \text{Tr} ({\bar{A_i}})(t))^2 \right\rangle \, .
\ee
Making use of the property (\ref{prop2}) it then follows that
\be \label{length}
l_{phys, i}(n_i) \, \sim \, n_i (for n_i < n_c) \,  .
\ee

Given the definition of coordinates and physical length it is possible to determine the diagonal components $g_{ii}$ of the spatial metric of the subspace of the three large dimensions via
\begin{equation}
g_{ii}(n_i)^{1/2} \, = \, \frac{d}{dn_i} l_{phys, i}(n_i) \, .
\end{equation} 
Making use of (\ref{length}) it then follows that $g_ii$ is independent of $n_i$:
\begin{equation}
g_{ii}(n_i,  t) \, = \, {\cal{A}}(t) \delta_{ii} \,\,\,   i = 1, 2, 3 \, .
\end{equation}
Invoking the $SO(3)$ symmetry of the subspace of the three large spatial dimensions it then follows that
\begin{equation} \label{metric}
g_{ij}({\bf n}, t) \, = \, {\cal{A}}(t) \delta_{ij} \,\,\,   i, j \, = \, 1, 2, 3 \, .
\end{equation}
The emergent metric is hence spatially flat, and the extent of the emergent space is infinite in the $N \rightarrow \infty$ limit. Note that ${\cal{A}}(t)$ plays the role of the square of the cosmological scale factor. Its time evolution at late times can be determined by solving the classical matrix model equations.  One finds \cite{IKKT-late}
\begin{equation}
{\cal{A}}(t) \, \sim \, t^{1/2} \, .
\end{equation}
This implies that there is no cosmological constant which appears in the evolution of the scale factor. This supports the view that the quantum aspect of the cosmological constant problem is an artefact of applying effective field theory in a context when it fails. In light of the discussion in Section \ref{section3} this is not surprising: the quantum theory is not based on quantizing any harmonic oscillators, and hence no quantum-induced cosmological constant is expected.

Note that it is only when ${\cal{A}}(t)$ becomes large that it makes sense to speak of a classical space. The space corresponding to the six small dimensions will never acquire a classical commutative limit. The classical space-time and metric obtained here are expectation values of matrix quantities when integrated over the set of matrices of the system. This is a difference compared to other approaches to obtain classical space-time from matrix models.
 
While the background in the matrix cosmology approach of \cite{BBL} is obtained from the $n = 0$ Matsubara modes, the $n \neq 0$ modes play an inportant role in determining the spectrum of cosmological perturbations. Since the starting point of the analysis is a thermal state of the BFSS matrix model, there are thermal fluctuations whose properties are determined by the partition function of the matrix model.  The computations \cite{BBL} follow what was done in the context of String Gas Cosmology. Specifically, given the matrix model partition function $Z(R, T)$ for a sub-volume of radius $R$, the internal energy is given by
\begin{equation}
E(R, T) \, = \, - \frac{\partial}{\partial \beta} {\rm{ln}} Z(\beta) \, ,
\end{equation}
where $\beta$ is the inverse temperature. The specific heat capacity $C_V(R, T)$ corresponding to this volume is then given by
\begin{equation}
C_V(R, T) \, = \, \frac{\partial}{\partial T} E(R, T) \, ,
\end{equation}   
from which the curvature fluctuations can be determined as detailed in (\ref{cosmopert}) and (\ref{spcheat}).The spectrum of gravitational waves can be determined by taking the partial derivative of the partition function with respect to $R$,  following what was done in String Gas Cosmology.

The resulting spectra of cosmological perturbations and gravitational waves are scale-invariant \cite{BBL}, as they are in the String Gas Cosmology model. The amplitude is of the same order of magnitude as what was obtained  in String Gas Cosmology. The spectrum of cosmological perturbations has a Poisson tail in the ultraviolet, as is to be expected in a system with thermal fluctuations.  The amplitude of the dimensionless power spectrum $P9k)$ of curvature fluctuations on IR scales is
\be
P(k) \, \sim \, \bigl( \frac{1}{l_s m_{pl}} \bigr)^2 \, .
\ee
The fact that the spectrum is scale-invariant on large scales is not surprising: since the matrix model is a non-perturbative definition of superstring theory, the thermodynamic quantities are expected to have the same holographic scaling as one obtains from the perturbative string partition function \cite{Tan, BV}. 

We have thus shown that the BFSS matrix model yields continuous emergent time running from $- \infty$ to $+ \infty$, continuous emergent three-dimensional space of infinite extent, and a spatially flat metric. There is no sign of any cosmological constant since matrix cosmology does not involve quantizing fields as sets of harmonic oscillators.   Thermal fluctuation yield scale-invariant spectra of curvature perturbations and gravitational waves. Hence, matrix cosmology is an alternative to cosmological inflation for solving the problems of Standard Big Bang Cosmology and for explaiining the origin of the observed structure in the Universe.  The cosmology based on the BFSS matrix model can be seen as an ultraviolet completion of String Gas Cosmology.

As mentioned before, in the approach summarized above,  classical space-time and the classical metric emerge as expectation values of matrix quantities. A very different approach to obtaining classical metric space-time from the IKKT matrix model has been put forwards and studiied in detail by Steinacker and his collaborators (see e.g. \cite{Stein1, Stein2} for reviews).  Here, it is proposed to start with a classical background space-time ${\cal{M}}$ and to determine an emergent metric in terms of a specific solution of the matrix model. Key to this approach is the establishment of a mapping between the matrix algebra  and the set of continuous functions on ${\cal{M}}$: Specifically, a matrix $\Phi$ gets mapped into a function $\phi$
\be
\Phi  \, \rightarrow \, \phi \in {\cal{C}}({\cal{M}}) \, ,
\ee
and the matrix commutator $[\Phi_1,  \Phi_2]$ is replaced by the Poisson bracket $\{\phi_1, \phi_2 \}$ (semi-classical limit).  Given a set of matrices $A^a$ which solve the IKKT matrix equations, it is then proposed to obtain a metric $G^{\mu \nu}$ on ${\cal{M}}$ through
\be
G^{\mu \nu} \, = \, \rho^{-2} \eta_{ab} E^{a\mu} E^{b\nu}
\ee
with
\be
E^{a\mu} \, = \, \{A^a, \gamma^{\mu} \} \, .
\ee

Using this correspondence, it has been possible to find solutions of the matrix model which in the semi-classical limit yield cosmological solutions, and in particular solutions corresponding to bouncing cosmologies. The matrix model action also yields the effective action for fluctuations which can be modelled as functions on ${\cal{M}}$.

For other work on emergent gravity from the IKKT matrix model the reader is referred to \cite{Klink}.

\subsection{Topological Phase}

In String Gas Cosmology, the evolution of the universe is divided into an early Hagedorn phase in which matter is a gas of closed strings in thermal equilibrium, and the decay of string winding modes triggers a phase transition to a phase in which three spatial dimensions expand as in the Standard Big Bang (SBB) model. Matrix cosmologies such as the one described in the previous subsection can be viewed as a realization of the early Hagedorn phase starting from basic principles of superstring theory. Another recent approach to determine the predictions of superstring theory for early universe cosmology is based on the idea that the early phase of the universe is a topological phase \cite{Vafa}.

Dualities (such as the T-duality symmetry) play a key role in the analysis of \cite{Vafa} (as it does in String Gas Cosmology \cite{BV}).  The usual degrees of freedom of matter which are excited in the current phase of cosmological expansion ({\it Phase II} in the notation of \cite{Vafa} become too heavy in the early universe ({\it Phase I} in the notation of \cite{Vafa}) and are frozen out. It is postulated that Phase I is a topological theory from the point of view of the variables in Phase II (i.e. those excited at late times). An example of such a theory is Witten's topological theory of \cite{Witten}.

Time is postulated to be common to both phases, and so is hence the notion of energy.  Since in Phase I no excitations involving momenta are excited the quantum state $ \vert \Psi >$ obeys
\be
P_i \vert \Psi > \, = \, 0 \, ,
\ee
and hence the background is homogeneous and isotropic, thus explaining the homogeneity problem of the SBB. Similarly, the local spatial curvature of this state vanishes since the curvature operator is related to the commutator of translation operators. Thus, the flatness problem of the SBB is addressed.

Since the topological theory is invariant of scale, the spectrum of cosmological fluctuations is predicted to be scale-invariant.  A solft breaking of the symmetry can be used to explain the observed slight red tilt of the spectrum. According to \cite{Vafa},  there are no primordial tensor modes in this scenario.

\subsection{De Sitter as a Coherent State}

There is known \cite{MN} that the ground state of supergravity cannot have a positive cosmological constant. It is either Minkowski space-time or anti-de-Sitter (AdS) space.  Over a time span of more than two decades there have been attempts to construct stable or metastable de Sitter states of string theory using effective field theory techniques with extra ingredients motivated by string theory.  One idea is {\it brane inflation} where inflation is generated by the potential energy of a moving brane in some bulk background geometry, the position of the brane in the bulk playing the role of the inflaton field (see e.g. \cite{Stephon, Tye, Burgess}).  One can also try to introduce D-branes to uplift the potental from an AdS minimum to a metastable de Sitter state (see e.g. \cite{KKLT}). Another attempt is to use the radius of one of the compatified dimensions of space as the inflaton. To obtain a sufficiently large period of inflation, the volume of the compactified space needs to be large, and hence this approach to string inflation is often called the {\it large volume scenario} \cite{large}. A further approach is to use monodromy to obtain a large field space which can yield inflation \cite{Eva}. For a review of these and other approached to string inflation see \cite{Baumann}.

The challenge for all of the above approaches to string inflation is to ensure that the constructions are not only consistent from the point of view of an effective field theory, but also from the higher dimensional string theory. Serious challenges have been raised.  Worries include singularities arising in the brane constructions \cite{Bena} and inconsistencies with the higher dimensional equations of motion \cite{Tatar} (see also \cite{Riet} for a general discussion). In addition,  most string inflation models violate the swampland criteria, and they are also in tension with the TCC.  In light of these concerns, it appears reasonable to expect that inflation can only (if at all) be realized from a non-perturbative and non EFT starting point.

These have recently been interesting attempts to construct cosmologies with de Sitter phases from non-perturbative constructions.  One approach is to construct a de Sitter phase as a coherent state of gravitons on top of a Minkowski background \cite{Dvali}. The claim is that such a state can be long-lived with a lifetime which is the {\it quantum break time} which is also the time scale when the back-reaction of infrared cosmological perturbations becomes important (see \cite{RHBbrReview} for an older review and \cite{IRback} for some newer references).  While the approach of \cite{Dvali} is not based on string theory, a recent approach of \cite{Keshav} is to construct a de Sitter phase as a coherent state in superstring theory, making use of key ingredients of M-theory \cite{Becker}. In this approach, the time scale of the de Sitter phase agrees with the TCC criterion.

\section{Challenges}
\label{section7}
 
 A nice feature of inflationary cosmology, our current paradigm of early universe cosmology, as formulated in the context of effective field theory (EFT),  is the fact that it admits a description which is self-consistent. However, as reviewed here, the usual EFT analysis of inflation is in tension with principles of quantum gravity and superstring theory.  Hence, if we want to truly understand the evolution of the early universe, we need to go beyond an EFT analysis. The challenge is that as soon as we go beyond the usual EFT framework, we lose perturbative controle. In the absence of such controle, it is crucial to focus on the effects of key symmetries and new degrees of freedom which differentiate superstring theory from the usual point particle-based EFTs.
 
 As has been reviewed here, cosmological inflation is not the only early universe scenario which can be consistent with current cosmological data. Bouncing and emergent cosmologies are interesting alternatives. In this article, a number of approaches towards obtaining bouncing and emergent cosmologies from superstring theory have been introduced.  Scenarios like Pre-Big-Bang cosmology,  Double Field Theory cosmology and to a certain extent also the Ekpyrotic scenario are based on effective field theory descriptions merged with key symmetries of string theory. These scenarios typically lead to bouncing cosmologies, but the challenge is that the EFT analysis breaks down near the bounce point. 
 
The study of truly non-perturbative approaches such as matrix model cosmology is at the very beginning. Although the initial results look promising, a major challenge is to make sure that the calculational schemes are under good controle. A long-standing challenge for superstring cosmology is to demonstrate that one can obtain both a successful cosmology and successful low energy particle physics phenomenology from a non-perturbative starting point for superstring theory.  An important issue is to address the mystery of Dark Energy. Based on the swampland and TCC criteria, Dark Energy cannot be a bare cosmological constant, and usual EFT descriptions are also constrained \cite{Lavinia}.  It is likely that the Dark Energy mystery will eventually lead us to very important insights about the nature of space, time and matter.

\begin{acknowledgements}

I acknowledge useful feedback on an initial draft of this article by Suddho Brahma, Keshav Dasgupta,  Heliudson Bernardo, Samel Laliberte and Cumrun Vafa, and I thank Anna Brandenberger for assistance with a couple of the figures.

I wish to thank the Pauli Center and the Institutes of Theoretical
Physics and of Particle- and Astrophysics of the ETH for
hospitality. The research of RB at McGill is supported in part by
funds from NSERC and from the Canada Research Chair program.    

\end{acknowledgements}


\end{document}